\documentstyle[referee,epsfig]{mn}
\def\etc{{etc.~\/}}
\def\eg{{e.g.\ }}
\def\etal{{ et al.~\/}}
\def\ie{{i.e.~\/}}
\def\mf{{Minkowski functionals }}
\def\dt{{\Delta T }}
\def\a{{\alpha }}
\def \aa    #1 #2   {{ A\&A \/}, {#1}, {#2}}
\def \aas   #1 #2   {{ A\&AS \/}, {#1}, {#2}}
\def \aj    #1 #2   {{ AJ \/}, {#1}, {#2}}
\def \apjl  #1 #2   {{ ApJ Lett. \/}, {#1}, {#2}}
\def \apj   #1 #2   {{ ApJ \/}, {#1}, {#2}}
\def \apjs  #1 #2   {{ ApJS \/}, {#1}, {#2}}
\def \mn    #1 #2   {{ MNRAS \/}, {#1}, {#2}}
\def \prl   #1 #2   {{ Phys. Rev. Lett. \/}, {#1}, {#2}}
\def \nat   #1 #2   {{ Nature \/}, {#1}, {#2}}
\def \com   #1 #2   {{ Comments Astrophys.\/}, {#1}, {#2}}
\def \sast  #1 #2   {{ Soviet Astron. \/} {#1}, {#2}}
\def \sastl #1 #2   {{ Soviet Astron. Lett.\/}, {#1}, {#2}}
\def \astr  #1 #2   {{ Astrophysics \/}, {#1}, {#2}}

\begin{document}
\title[Morphological non-Gaussianity]
{Testing non-Gaussianity in CMB maps by
morphological statistic}
\author[S.Shandarin]
{Sergei F. Shandarin \\
The University of Kansas, Department of Physics and Astronomy,
1082 Malott Hall, 1251 Wesco Hall Drive,\\
Lawrence, KS 66045, U.S.A.}
\maketitle
\begin{abstract} 
The assumption of Gaussianity of the primordial perturbations plays 
an important role in modern cosmology. The most direct test of this
hypothesis consists in testing Gaussianity of the CMB maps. 
Counting the pixels with the temperatures in
given ranges and thus estimating the one point probability function
of the field is the simplest of all the tests.
Other usually more complex tests of Gaussianity generally 
use a great deal of the 
information already contained in the probability function. However, 
the most interesting outcome of such a test 
would be the signal of non-Gaussianity independent 
of the probability function. 
It is shown that the independent information 
has purely morphological character \ie it depends
on the geometry and topology of the level contours only.
As an example we discuss in detail the quadratic
model $v=u+\alpha (u^2-1)$ ($u$ is a Gaussian field with $\bar{u}=0$ and
$<u^2>=1$, $\alpha$ is a parameter) 
which may arise in slow-roll or two-field 
inflation models. 
We show that in the limit of small amplitude $\alpha$
the full information about the non-Gaussianity
is contained in the probability function. If other tests 
are performed on this model they simply recycle the same 
information. 
A simple procedure allowing to assess
the sensitivity of any statistics to the morphological information
is suggested.  
We provide 
an analytic estimate of the statistical limit for detecting the quadratic
non-Gaussianity $\a_c$ as a function of the map size in the ideal situation 
when the scale of the field is resolved. This  estimate
is in a good agreement with the results of the Monte Carlo simulations
of $256^2$ and $1024^2$ maps. The 
effect of resolution on the detection quadratic non-Gaussianity
is also briefly discussed. 
\end{abstract}

\section{Introduction}
The quest for the physical mechanism of the generation of the
initial inhomogeneities 
along with the measurements of the major cosmological parameters 
($H_0, \Omega_{\Lambda}, \Omega_{CDM}, \Omega_b, P(k),$ etc) 
is one of the most important problems in modern cosmology. 
The standard inflationary model predicts
the primordial fluctuations
were Gaussian random fields
\cite{gut-pi82,haw82,sta82,bar-ste-tur83}. 
In agreement with the theory the current observations 
provide little evidence for deviations from Gaussianity.
The majority of the tests of Gaussianity in the COBE maps
\cite{col-etal96,kog-etal96,fer-mag-gor98,nov-fel-sh99,pan-val-fan98,bro-teg99,ban-zar-gor00,muk-hob-las00,bar-etal00,agh-for-bou01,phi-kog01}
have resulted  in the general agreement that all non-Gaussian signals
were of noncosmological origin 
\footnote{However, Magueijo \shortcite{mag00} still has a 97\%
confidence level that the signal is not due to systematics.}.
This was not perhaps a surprise 
because of a very large physical 
scale corresponding to the COBE resolution ($\approx 7$ deg).
Recent studies of the $\dt/T$ maps on a degree and sub-degree scales also 
showed no significant deviations from Gaussianity 
\cite{par-etal01,sh-etal01,wu-etal01}.  
 
Nevertheless,
the question of possible non-Gaussianity in the CMB maps is very important
because of the following reasons.
First,  a detection of a non-Gaussian component in the primordial
fluctuations may profoundly affect modern cosmology ruling out 
some models of the early universe and boosting the others 
(see \eg Turner \shortcite{tur97}; 
Vilenkin \& Shellard \shortcite{vil-she94}). Second,
Gaussianity is a key underlying assumption of all experimental 
power spectrum analyses to date, entering into the computation of 
error bars \cite{teg97,bon-jaf98}, and therefore needs to be 
observationally tested. In addition, the hypothesis of the
Gaussianity of the initial perturbations
enters in the studies of the large-scale structure, clusters of galaxies,
formation of galaxies, $Lya$ forest, etc.
Finally, studying Gaussianity of CMB maps 
may reveal otherwise undetected foreground contamination.


Many tests for non-Gaussianity have been suggested.
For instance, in the recent paper by Wu \etal \shortcite{wu-etal01} the authors
employed a total 82 hypothesis tests for Gaussianity, although,
as the authors noted, the tests were not independent.
Obviously the question arises how independent 
the different tests were.
A related issue concerns the possibility of constructing a set of
independent tests. The full solution to this problem is 
beyond the scope of this paper but we describe a simple procedure
that allows to assess the independency of a test from the probability 
function. 
The probability function, $f(u)$ also called the probability density function
tells what is the probability that a randomly chosen point of the field,
$u(x,y)$ has a certain value. 
In practice $f(u)du$ is often estimated by counting the cells having
the values of the field between $u$ and $u+du$. 
Gaussian fields have the Gaussian probability function. 
The choice of the probability function (PF)
as the reference statistic is due to both  
its conceptual and practical simplicity.
Generally speaking
the more sensitive a test to the statistical information different from
the probability function the more useful it is.

The hierarchy of n-point correlation functions or equivalently n-spectra
is also widely used for testing non-Gaussianity.
However, compared to PF this is not as easy task from both 
conceptual and practical points of view.
One reason is the multidimensional character of the n-point functions.
The n-point function as well as n-spectrum in the two-dimensional space
is a function of $2n-3$ variables. Thus, testing the lowest order functions
sensitive to non-Gaussianity, three-point correlation function or
bispectrum, one has to deal with a function of three variables.
So far pragmatic solutions to this problem were either 
computing a small number of particular cuts 
in the three-dimensional space or introducing some integral quantities.
Both shortcuts obviously result in incompleteness of the test.
The other reason is purely computational: 
computing of a n-point function on the grid with $N_{pix}$
pixels using current methods requires $O(N_{p}^n)$ operations 
which is already prohibitive for current fairly small maps 
(COBE, QMASK, Maxima I) even for $n=4-5$. 
Although, a clever technique using kd-trees can potentially reduce it to 
$O(N_{pix}(logN_{pix})^{n-1})$ \cite{szap-etal00} it has to be developed yet. 

The n-point functions carry information about 
the maps in highly redundant and diluted form. 
In order to see this let us consider a large two-dimensional map
obtained observationally or from a theoretical simulation.
Obviously all information about the map can be stored in the form of a
function of two variables (\eg the map itself). 
The two-point function of the map is a function of one variable and
thus considerably reduces the information about the map by loosing
the phase information. 
In general case the three-point function also 
considerably reduces the information about the map but
in contrast to the two-point function it increases the dimensions of the
space that means a huge dilution of information. 
The four-point
function in the two dimensional space is a function of five
variables meaning that it dilutes the information 
even more than the three-point function. In general case, 
the higher order of the n-point function 
the more diluted is the information about the field.

It is well known that every n-point function affects the 
probability function, see \eg White (1979) and Balian and Schaeffer (1989).
It means that the PF carries some information about every 
n-point function. Reversing this statement one can
say that no n-point function carries information that is 
completely independent from the PF.
Thus, one may also ask
what information is stored in n-point functions which
is independent of the PF and whether it is possible to extract it
or at least to assess it. 
Obviously, the same question must be addressed not to only
the n-point functions but to all other statistics. 
These issues are discussed bellow.

The PF  
or equivalently the cumulative probability function (CPF)
\footnote{Here we would like to emphasize the informational content 
of a non-Gaussian statistic and therefore do not distinguish the 
PF and CPF assuming that both contain the same information.}   
is not only the simplest conceptually but also most efficient numerically. 
Computing this statistic requires only $O(N_{pix})$ operations. 
The only problem is that the Gaussian PF does not guarantee the Gaussianity
of the field. 
Therefore, some additional statistical information is badly needed
in case when the PF of the field is Gaussian since if the PF is
non-Gaussian the non-Gaussianity is already detected. 
The next step obviously would be the identification of the
physical process responsible for the non-Gaussianity but
first it must be detected.
Thus, if the PF is Gaussian the additional 
information must be independent of that contained in the PF.
We will show that such information has purely morphological character. 
This means that it is completely determined by 
the geometric and topological statistic of the excursion sets.
Thus, a set of morphological parameters based on Minkowski functionals
becomes a natural choice of the statistics that is 
sensitive to non-Gaussianity and completely independent of the PF
provided that proper parameterization is used.

A particular kind of non-Gaussianity known as the quadratic model 
has recently attracted much attention \cite{col-bar87,luo-sch93,mat-ver-jim00,ver-etal00,ver01,ver-etal01}.
One reason is that it
could be generated by plausible physical mechanisms in the early 
universe \cite{fal-ran-sre93,gan-etal94,luo94}. 
The other is a relative ease of its analysis. In this
paper we show that the simplest test for Gaussianity,
the probability function, provides also the complete
statistical information in the most interesting case of small
amplitudes. It means that other tests if applied to this model at best only
recycle a part (probably small) of this non-Gaussian information.
In the general case of arbitrary amplitude 
the set of global \mf completely characterize the statistical
properties of this field. 

The rest of the paper is organized as follows. We describe the set
of morphological quantities in Sec.2. Sec.3 describes a particular
parameterization of the morphological statistics that makes them
PF-independent. As an illustration we discuss a couple of simple non-Gaussian
models one of which is the quadratic model that is often used in
cosmology in Sec.4. Then, in Sec.5 we describe a class of the
simplest non-Gaussian fields which  can be called trivial non-Gaussian
fields. 
Section 6 describes a few simple estimators of 
the amplitude of the quadratic non-Gaussianity. We describe simple
Monte Carlo simulations modeling the detection of the quadratic
non-Gaussianity in Sec. 7.
Finally, we discuss the results in Sec. 8.
 
\section{Morphological Quantities in Two Dimensions}

Morphology of two-dimensional random fields can be conveniently
described in terms of geometric and topological properties of the 
regions bounded by the contours of constant level. There is a
particularly useful set of quantities called Minkowski functionals 
\cite{min03} which have very simple geometric and topological
interpretations. 
For each isolated region bounded by a contour there are only three scalar 
Minkowski functionals: the area within the boundary, $a_i$, 
its perimeter or the contour length, $c_i$, 
and the Euler characteristic or  genus, $g_i$
which is $1-n_{hi}$ where $n_{hi}$ is the number of holes in
the region.
 
Minkowski functionals are additive quantities and therefore 
can be easily calculated for any set of regions if they are known for
each region. In particular,
the global Minkowski functionals, i.e. the total area, $A$, 
contour length, $C$ and genus, $G$ of the excursion set:
\begin{equation}
A=\sum_i a_i~,\qquad C=\sum_i c_i~,\qquad G=\sum_i g_i
\end{equation}
are often used 
\cite{got-etal90,win-kos97,sch-gor98,nov-fel-sh99,nov-sch-muk00,wu-etal01,sh-etal01}.

The total area of the excursion set, $A$ is obviously the CPF of the field: 
$A(u) \equiv F(u)=\int_u^{\infty} f(u')du'$.  
The Euler characteristic or 
genus have been used in cosmology for a number of years
\cite{dor70,got-etal86,col88}.

The first time the set of global Minkowski functionals was introduced 
into cosmology with the reference to their significance in differential and
integral geometry by  Mecke, Buchert \& Wagner \shortcite{mec-buc-wag94}
and Schmalzing \& Buchert \shortcite{sch-buc97}. 
In particular, they emphasized 
a powerful theorem by Hadwiger \shortcite{had57}
stating that under rather broad restrictions the set of \mf 
provides a complete description of the morphology (for further discussion
see e.g. Kerscher \shortcite{ker99}).

In addition, the Minkowski functionals of the largest (by area) region
($A_p$, $C_p$, and $G_p$)
give accurate description of the percolation phase transition 
\cite{yes-sh96}. At percolation
the regions merge into one region that spans throughout the whole
space of the field. Percolation phase transition is sensitive to
some types on non-Gaussianity \cite{zel82,sh83,sh-zel83,sh-zel84}.

\section{Parameterization}
Often the level of the excursion set, $u$ is used 
to parameterize Minkowski functionals.
The global \mf of a Gaussian field as functions of the level
are given by simple analytic formulae \cite{lon-hig57,got-etal90}
\begin{eqnarray}
A(u)&=&\frac{1}{2} \left[1-{\rm erf}\left(\frac{u}{\sqrt{2}}\right)\right], \cr
C(u)&=&\frac{1}{2R}\exp\left(-\frac{u^2}{2}\right),\cr
G(u)&=&\frac{1}{(2\pi)^\frac{3}{2}} \frac{1}{R^2}u 
\exp\left(-\frac{u^2}{2}\right), \label{GMF_G}
\end{eqnarray}
where $R=\sqrt{2}/\sigma_1$ is the scale of the field;
$\sigma_1$ is the rms of the first derivatives 
(in statistically isotropic fields both derivatives 
$\partial u/\partial x$ and $\partial u/\partial y$ have equal rms). 
It is assumed that the field is normalized: $<u>=0$ and $<u^2>=1$.
The \mf as functions of the level are shown in the left hand side 
panels of Fig.1.

The  parameterization by the level is useful for many applications. 
However, for the study of the morphology of the fields
and Gaussianity in particular
the parameterization by $A$ is much better because it makes the 
statistics independent of the PF and considerably less correlated 
with each other  \cite{sh-etal01}.

The total area of the excursion set $A$ of the field 
was suggested to parameterize other quantities \cite{yes-sh96,sh-etal01}. 
Parameter $A$ is a single valued function of ${\nu}$, the parameter used 
in the most papers studied genus 
(Park \etal \shortcite{par-etal01}  and references therein) 
and therefore every function of ${\nu}$ can be also expressed
as a function of $A$.
However  $A$ is more directly related to the morphology of random fields. 
In addition, being equal to the cumulative probability function
(CPF) it has a very simple statistical and geometrical meaning. 
Imagine that the excursion set ($u>u_c$) is painted black 
while the rest of the map ($u<u_c$) remains white.  
Then the fraction of the area in black equals $A$.
The right hand side panels in Fig.1 show the level, $u$, 
total contour length, $C$ and genus, $G$ as a function 
of the total area of the excursion set $A$ for a Gaussian field. 

It has been noticed that percolation properties of the field can
be useful for detecting some types of non-Gaussianity 
\cite{zel82,sh83,sh-zel83,sh-zel84,kly-sh93}. 
The \mf of the largest by area region are
excellent parameters to characterize the percolation properties of
the field. In contrast to the global \mf they are not known in an analytic 
form. Figure 2 shows the \mf of the largest region for a Gaussian 
field obtained from a large ensemble ($N=1000$) of large Monte Carlo 
simulations ($1024^2$ grid).

It should be stressed that the major reason of using $A$ as an independent 
parameter consists in isolating independent morphological information that
is not present in the probability function (PF). For example, $C=C(A)$ 
carries only PF-independent information, while $C=C(u)$ mixes it up
with the information stored in the PF.

\section{Examples of Non-Gaussian Fields}
It is interesting to compare how some of the n-point functions 
and morphological
characteristics signal the presence of non-Gaussianity 
in the field. We consider two examples with quadratic and cubic
non-Gaussianity. The former has been suggested as a model having
plausible physical mechanisms producing small deviations from
Gaussianity \cite{fal-ran-sre93,gan-etal94,luo94}. 
The latter has no physical motivations and is taken as a toy model only.

\subsection{Cubic Model}
First we consider a transformation 
\begin{equation}
v=u+\beta u^3,
\end{equation}
where $\beta$ is assumed to be positive, which guarantees monotonicity
of the mapping. The parent field $u$ is assumed to be a normalized Gaussian
field with $<u>=0$ and $<u^2>=1$.
The CPF of the resulting  field is
\begin{equation}
F(v)=F_G(u_s),
\end{equation}
where $u_s$ is the solution of the cubic equation $u_s+\beta u_s^3=v$
\footnote{For the positive $\beta$ there is only one real solution $u_s$.}.
Although the solution could be obtained in a closed form the expression 
is quite cumbersome therefore we take the limiting case of small $\beta$
\begin{equation}
u_s\approx v-\frac{\beta v^3}{1+3\beta v^2}.
\end{equation} 
Thus, the CPF becomes
\begin{equation}
F(v)=\frac{1}{2}\left\{1-{\rm erf}\left[\frac{1}{\sqrt{2}}\left(v-\frac{\beta v^3}{1+3\beta v^2}\right)\right]\right\}
\end{equation}
and the PF can be obtained by differentiation: $f(v)=-dF/dv$.

The two- and three-point correlation functions of this field can 
be easily obtained for an arbitrary $\beta$
\begin{eqnarray}
\xi_2(1,2)&=&(1+6\beta)\xi_g(1,2)+ 3\beta^2\xi_g(1,2)\left[3+2\xi_g^2(1,2)\right], \cr
\xi_3(1,2,3)&=&0
\end{eqnarray}
where $\xi_i(1,2)$ is a shortcut to $\xi_i(r_1,r_2)$. The function $\xi_g$
is the two-point correlation function of the parent field $u$.
To the linear order in $\beta$
the two point function remain the same by the form but 
acquires a different normalization $(1+6\beta)$.
The reason why the three point function does not show the presence of
non-Gaussianity is the symmetry of the mapping: $u \rightarrow v$
is mapped by an odd function. 

Global Minkowski functionals $C(A)$ and $G(A)$ 
as well as the percolation statistics $A_p=A_p(A)$, $C_p=C_p(A)$ 
and $G_p=G_p(A)$
remain exactly same as in the parent Gaussian field since the transformation
simply relabels the levels without changing the contour lines. 
However, if they were expressed as functions of
the level ($C(v)$ and $G(v)$) they would differ from  $C_G$ and $G_G$ 
because some non-Gaussian signal leaks into them from the CPF.
We will illustrate this point when discuss the quadratic model.

The full information about non-Gaussianity of the field 
is obviously stored in one number (the value of $\beta$)
and can be obtained from one point statistics (the 
PF or equivalently CPF) however the non-Gaussianity is not detected by
the odd order n-point statistics.
In a generic case of monotonic mapping
$v=\phi(u)$ where $\phi$ is a monotonic but not necessarily odd function 
of $u$ all n-point functions would detect non-Gaussianity.

\subsection{Quadratic Model}
As it was mentioned before the quadratic non-Gaussianity
is particularly popular in cosmology.
The quadratic model is the sum of a Gaussian field and its square
\begin{equation}
v=u+\alpha (u^2-1). \label{GaG2}
\end{equation} 
In this paper we assume that $u$ is normalized to unity: $<u>=0$ and $<u^2>=1$.
If one likes a different normalization $<u'>=0$ but $<u'^2>=\sigma'^2$
and the quadratic transformation in the form
\begin{equation}
v'=u'+\alpha' (u'^2-\sigma'^2)
\end{equation}
then the relations between the parameters of the two transformations 
are as follows
\begin{equation}
u'=\sigma'u,~~~v'=\sigma'v,~~~\alpha'=\alpha/\sigma'.
\end{equation}
For certainty, without losing generality we will assume $\alpha >0$. 
Solving eq. \ref{GaG2} for $u$ and denoting the solutions
as $u_1$ and $u_2$  ($u_2 < u_1$) one obtains 
\begin{equation}
u_1=\frac{\sqrt{1+4\alpha(\alpha+v)}-1}{2\sqrt{2}\alpha},~~~~~
u_2=\frac{-\sqrt{1+4\alpha(\alpha+v)}-1}{2\sqrt{2}\alpha}. \label{solG2}
\end{equation}
The CPF of the field can be written then as
\begin{eqnarray}
F(v) &=&1-\frac{1}{2}\left[{\rm erf}\left(
\frac{\sqrt{1+4\alpha(\alpha+v)}-1}{2\sqrt{2}\alpha}\right) 
- {\rm erf}\left(\frac{-\sqrt{1+4\alpha(\alpha+v)}-1}{2\sqrt{2}\alpha}\right)\right]. \label{CPF_G2}
\end{eqnarray} 
Differentiating it with respect to $v$ one easily obtains the PF
\cite{luo-sch93,mat-ver-jim00}
\footnote{In the paper by Luo \& Schramm \shortcite{luo-sch93} 
the second term 
is missed which is not important for small $\alpha$.}:
\begin{equation}
f(v) = \frac{1}{\sqrt{2\pi}} 
\frac{\exp \left(-\frac{(\sqrt{1+4\alpha(\alpha+v)}-1)^2}
{8\alpha^2}\right)
 + \exp \left(-\frac{(-\sqrt{1+4\alpha(\alpha+v)}-1)^2}{8\alpha^2}
\right) }{\sqrt{1+4\alpha(\alpha+v)}}. \label{pf_g2}
\end{equation}

The PF has a weak singularity at $v=-1/(4\alpha) -\alpha$
\begin{equation}
f(v) \approx \sqrt{ \frac{2 }{ \pi}} \frac{\exp(-\frac{1}{8\alpha^2})}
{\sqrt{1+4\alpha(\alpha+v)}}. \label{asymp}
\end{equation}
The PF is shown in Fig.3 for a several values of the parameter $\alpha$.
The left hand side panel shows the PF for a relatively large amplitudes 
$\a=0.25,~ 0.18,~0.1, ~ 0.05$.
For small amplitudes a better illustration is the difference
of the PF and Gaussian PF shown in the right hand side panel
($\a=0.02,~ 0.01, ~0.005$).
Note that for small $\a$ 
\begin{equation}
f(v) \approx \left[1+\a h(v)\right]f_G(v) \label{pf_gag2_lim}
\end{equation}
where 
\begin{equation}
h(v) =  v (v^2-3)
\end{equation}
and $f_G(v)$ is the Gaussian PF. The function 
$0.02h(v)f_G(v)$ is shown by the solid line in the right hand side
panel of Fig. 3. It almost merge with the exact PF shown by the dotted
line. The agreement is even better for smaller values of $\a$.

The two and three point functions are respectively
\begin{eqnarray}
\xi_2(1,2)&=&\xi_g(1,2)+ 2\alpha^2\xi_g^2(1,2), \cr
\xi_3(1,2,3)&=&2\alpha\left[\xi_g(1,2)\xi_g(1,3)+\xi_g(1,2)\xi_g(2,3)+\xi_g(1,3)\xi_g(2,3)\right] \cr
&+&8\alpha^3\xi_g(1,2)\xi_g(1,3)\xi_g(2,3) \label{gag2_3pf}.
\end{eqnarray}
In the case of quadratic mapping (\ref{GaG2}) the three-point function
detects non-Gaussianity in the linear order in $\a$. 
In this case all even order functions vanish in linear order of $\alpha$
due to symmetry of the mapping.

The global Minkowski functionals of the field can be also readily obtained 
\begin{eqnarray}
C(v)&=&C_G\left(\frac{\sqrt{1+4\alpha(\alpha+v)}-1}{2\sqrt{2}\alpha}\right) 
+C_G\left(\frac{-\sqrt{1+4\alpha(\alpha+v)}-1}{2\sqrt{2}\alpha}\right) \cr
G(v)&=&G_G\left(\frac{\sqrt{1+4\alpha(\alpha+v)}-1}{2\sqrt{2}\alpha}\right)
-G_G\left(\frac{-\sqrt{1+4\alpha(\alpha+v)}-1}{2\sqrt{2}\alpha}\right)
\label{GMF_G2}
\end{eqnarray}
where $C_G(u)$ and $G_G(u)$ are the Gaussian global Minkowski functionals
(eq.\ref{GMF_G}). The $\chi^2$ field with one degree of freedom 
is obviously a particular limiting case of this model and has been 
discussed in great detail \cite{tom90,wor94,col-bar87,sch99,nov-sch-muk00}. 

Figure 4 (left hand side panels) shows global \mf as
a function of the level for four values of the amplitude $\a =0.25,~ 0.18,
~0.1, ~ 0.05$.
Combining the equation for the CPF (\ref{CPF_G2}) and
recalling that $A(v)\equiv F(v)$ one can also  plot $C$ and $G$
as functions of $A$  (Fig.4,  right hand side panels). 
The Gaussian curves are shown by solid lines in all panels.
One easily sees that the non-Gaussian signature is much stronger 
in the left hand side panels. This is due to leaking of some non-Gaussian
information into the $C(v)$ and $G(v)$ curves from the PF. 
The curves $C(A)$ and $G(A)$
show only the non-Gaussianity that is absent in PF and therefore show nothing
when such information is absent. This is why one can see only two non-Gaussian
curves corresponding to $\a=0.25$ and $0.18$ in the right hand side panels.
Although all four non-Gaussian curves are plotted
the curves corresponding to $\a=0.1$ and $0.05$ are merged with the
Gaussian curves and are not seen. 
The curves corresponding to $\a=0.1$ and $0.05$ are clearly
seen in the left hand side panels
although, as shown below, the morphology of these fields is practically 
Gaussian  or more exactly the non-Gaussian morphology is 
completely absent in maps with $N_{pix} < 10^7$. 
Another illustration of this effect
is provided by the comparison of the $C(v)$ and $G(v)$ curves at $v>1$
with the $C(A)$ and $G(A)$ at $A<0.16$. The mapping 
(eq. \ref{GaG2}) is monotonic and thus does not change the morphology
of the field at small amplitudes in this range of $v$ or $A$. 
All the curves
merge in the right hand side panels manifesting the similarity of
the morphology to the Gaussian one, but they are clearly distinct
in the left hand side panels due to leaking of some non-Gaussianity from 
the PF.

As a whole the quadratic model (\ref{GaG2}) is not a monotonic mapping:
$v$ is monotonically increasing at at $u > -1/2\alpha$
and monotonically decreasing at $u > -1/2\alpha$. 
This results in two terms in eq. \ref{CPF_G2} and eq. \ref{GMF_G2}
as well as in the deviation of the morphology from the Gaussian one
because the regions belonging to the excursion set $v > v_c$ 
belong to different excursion sets in the parent field $u$:
$S_1~ [u>u_1(v_c)]$ and $S_2~ [u<u_2(v_c)]$ (eq. \ref{solG2}).

Consider now the limiting case of small $\alpha$. For small
$\alpha$ the decreasing branch of the mapping is shifted to
large negative $u$ and thus greatly reducing the second terms in 
eq. (\ref{pf_g2} and \ref{GMF_G2}). Therefore,
\begin{eqnarray}
A(v) &\approx&\frac{1}{2}\left[1-{\rm erf}\left(
\frac{\sqrt{1+4\alpha(\alpha+v)}-1}{2\sqrt{2}\alpha}\right)\right] \cr
C(v)&\approx&C_G\left(\frac{\sqrt{1+4\alpha(\alpha+v)}-1}{2\sqrt{2}\alpha}\right)\cr 
G(v)&\approx&G_G\left(\frac{\sqrt{1+4\alpha(\alpha+v)}-1}{2\sqrt{2}\alpha}\right)
\end{eqnarray}
and the mapping becomes almost monotonic.

In cosmology one is usually interested in detecting the possibly smallest
non-Gaussianity. For small $\a$ the quadratic
transformation can be practically monotonic for a map of a finite size.
Let us consider a map with $N_{pix}$ pixels and
calculate the critical value $\alpha_c$
corresponding to the value of $\alpha$ when the double valued character
of transformation (\ref{GaG2}) is observed on average 
at one pixel only.
The minimum of the parabola (\ref{GaG2}) is reached at $u_{min}=-1/2\alpha$.
Therefore the probability that a map of the size $N_{pix}$ 
contains on average
one pixel where the monotonicity of (\ref{GaG2}) is broken is
\begin{equation}
1-F_G(u_{min})\equiv 1-F_G(-1/2\a_c)=\frac{1}{N_{pix}}.
\end{equation}
The solution for $N_{pix}$ has a simple analytic form  for $\a_c \ll1$ 
\begin{equation}
N_{pix}\approx \sqrt{\frac{\pi}{2}}\frac{1}{\a_c}\exp(\frac{1}{8\a_c^2}).
\label{alpha_crit}
\end{equation}
Critical amplitude $\a_c$ is plotted as a function of $\log_{10}(N_{pix})$
in Fig.5 with the points corresponding to the best resolutions of the COBE,
MAP and Plank experiments. One easily sees that only relatively large $\alpha$ 
(roughly $\a> \a_c \approx 0.09$)  may result in 
the deviation of the quadratic model (\ref{GaG2}) from monotonicity
for even very large maps. It means that the full information about
the non-Gaussianity of the quadratic model can be obtained from a one 
point function only if $\a < 0.09$.
Nothing can be gained by applying more complex n-point statistics.

\section{Trivial Non-Gaussianity}

From the examples of Sec. 4 one can make a further step 
to a little more general case.
Consider a monotonic mapping of a Gaussian field $u \rightarrow v$:
$v=\phi(u)$ with $dv/du > 0$ (the case with  $dv/du < 0$ is similar).
This transformation obviously affect only the PF
\begin{equation}
F(v)=F_G(\phi^{-1}(v))
\end{equation}
where $F_G$ is the Gaussian cumulative probability function
and $\phi^{-1}(v)$ is the inverse of the function $\phi(u)$.
The PF
can be found by the  differentiation of the above expression.

The shapes of the level contours do not change in the resulting field
because the contours of constant $v$ coincide with  
the corresponding contours of constant $u$.
The non-Gaussian fields obtained by this kind of mapping can
be called trivial since all its non-Gaussianity is described by
the one-point function (\eg PF).

The cubic as well as quadratic model with roughly 
$\alpha < 0.09$ (in maps with $N_{pix} < 10^7$) 
are examples of trivial non-Gaussian fields.
Another example of a trivial non-Gaussian field  is given by 
the exponential mapping
\begin{equation}
v=\exp(u)
\end{equation}
where $u$ is a Gaussian field. The resulting field has the lognormal PF and 
exactly the Gaussian morphology. 

If $v=\phi(u)$ is not a monotonic function then $u=\phi^{-1}(v)$
is a multiple valued function.
As a result some regions of the excursion set with $u > u_c$ may not 
satisfy the condition $v(u) > v_c(u_c)$ and vice versa. This case is
a little more complex but also can be fully treated analytically
in terms of the \mf.

One point function PF (or CPF) is in many respects the simplest statistics
requiring only $O(N_{pix})$ operations. 
The other statistics (e.g. n-point correlation functions) share some
information with the PF. The morphological statistics (\mf)
are easy to isolate from the PF by parameterizing them by the area
of the excursion set. We are not aware of the general method
of isolation the morphological non-Gaussianity in other
non-Gaussian statistics. However, there is a simple technique
allowing to assess the sensitivity of any statistics to the morphological
non-Gaussianity. 
 
Let us relabel the levels of the field $v \rightarrow u$ according to
the equation
\begin{equation}
F(v) =F_G(u) \label{gausssianization}
\end{equation}
where $F(v)$ and $F_G(u)$ are the CPF of the original and Gaussian
fields respectively. The new field has the  Gaussian PF
by design but restores the morphology of the original field
because the contour lines have been only relabeled but not distorted. 
Then computing
any non-Gaussian statistics for the new field one can be sure
that it
depends only on the information that is absent in the PF. 
For instance, if the original field was trivial non-Gaussian then after
this transformation all non-Gaussian statistics must vanish.
This transformation of the field can be called the gaussianization 
of the field. It was used for the purpose of recovery of the
primordial fluctuations from the present day galaxy distribution 
\cite{wei92} and recovery of the power spectrum from QSO  $Ly\alpha$
forest spectra \cite{cro-etal99}.

It is worth stressing that the gaussianization neither simply smoothes 
the original field no re-scales the spectrum as in the transformation  
suggested by Wu \shortcite{wu01}. 
It is local but in general highly nonlinear and
its only purpose is to isolate non-Gaussian information which is
complimentary to the PF \cite{sh-etal01}. 
It certainly affect all n-point functions
and n-point spectra including the spectrum of the field. It also 
changes the phases and therefore may be useful in studies similar
to that by Coles \& Chiang \shortcite{col-chi00}.

\section{The Amplitude of the Quadratic model} 

Assuming that detecting the possibly smallest non-Gaussianity is the
goal we consider a few estimators of the amplitude of the quadratic
non-Gaussianity $\a$ ignoring minute effects on morphology. It means that
$\a$ is to be less than $\a_c$ of eq. \ref{alpha_crit} (see
also Fig. 5).
Luo and Schramm (1993) derived the skewness in the quadratic model. 
To linear approximation in $\alpha$ it is
\begin{equation}
S \equiv \frac{<(v-\bar{v})^3>}{\sigma_v^3} = <v^3>\approx 6\a.\label{a_3}
\end{equation}
The last equality is because $\bar{v}=0$ and $\sigma_v=1$.
Thus, measuring the skewness one can estimate the amplitude of the
quadratic component. One can also easily obtain the estimates 
of $\alpha$ from higher central moments
\begin{equation}
c_5\approx 60\alpha, ~~~~c_7\approx 630\alpha \label{a_5_7} 
\end{equation}
where $c_i$ are the corresponding central moments 
$c_i=<v^i>$ since $<v>=0$ and $<v^2>=1$. 
In principle, any of odd 
moments can be used  for the estimate of the amplitude $\alpha$.

Since the higher moments
depend stronger on the tails of the PF one might think that the most 
accurate is the estimate based on the skewness. As we shall see 
this trend  is very week at least in the case of the lowest three moments 
(n=3, 5 and 7). 

Using the asymptotic form of $f(v)$ (eq. \ref{asymp}) one also can derive 
$\a$ using the least square estimator. 
Suppose there are $N$ measurements of $v$ binned in $N_b$ equal
bins, $\delta v$, ($n_i, ~~ i=1,\cdots,N_b$). Then minimizing the $\chi^2$ 
\begin{equation} 
\chi^2 =\sum\limits_{i=1}^{N_b}\left[\frac{n_i-n(\a,v_i))^2}{\sigma_i}\right] 
\end{equation}
one can obtain $\a$. Simple calculations result in
\begin{equation} 
\a \approx \left\{\sum\limits_{i=1}^{N_b}\frac{\left[ n_i^2-N^2f_G(v_i)^2 \delta v^2\right]h(v_i)}{Nf_G(v_i)\delta v}\right\}
\left\{2\sum\limits_{i=1}^{N_b}\frac{n_i^2h^2(v_i)}{Nf_G(v_i)\delta v}\right\}^{-1}. \label{a_chi}
\end{equation}
Considering the propagation of errors one can also estimate the standard 
deviation of $\a$
\begin{equation}
\sigma_{\a} \approx \left\{2\sum\limits_{i=1}^{N_b}\frac{n_i^2h^2(v_i)}{Nf_G(v_i)\delta v}\right\}^{-1/2}. \label{sigma_a}
\end{equation}
Assuming $n_i \gg 1$ one can simplify eq.\ref{sigma_a}
\begin{equation}
\sigma_{\a} \approx N^{-1/2} \left(\sum\limits_{i=1}^{N_b} f_G(v_i)h^2(v_i) \delta v\right)^{-1/2}  
\approx N^{-1/2} \left(\int\limits_{-\infty}^{\infty} f_G(v)h^2(v) d v\right)^{-1/2} = (6N)^{-1/2}. \label{limit_a}
\end{equation}
This can be used for the estimate of the statistical limit
on the value of $\a$ on a grid of size $N_{pix}$. Requiring that
$|\a|>2\sigma_{\a}$ ($2\sigma$ detection) one obtains
\begin{equation}
|\a| > (3N_{pix}/2)^{-1/2}.  \label{2sig_det}
\end{equation} 
Thus, for the map of the size of $N_{pix} \sim 10^6$ the value $\a$ must be
greater than $\a_{min} \sim 10^{-3}$ in order to be detected by
this method. For an actual CMB experiment eq. \ref{limit_a} is good but
still an approximation. The effects of the radiation transfer function
are needed to be included in a realistic application to CMB map.

\section{Monte Carlo Simulations}
In order to check the above analytic calculations  
and study the other effects we use simple Monte Carlo 
simulations. First, a Gaussian field $u$ ($<u>=0, <u^2>=1$) is generated 
in a square of the size $N_{pix}=N_1\times N_1$. We used the flat 
spectrum smoothed at $k_s=N_1$ in units of the fundamental
wavelength, $P \propto \exp(-k^2/k_s^2)$. Then, the field $v=u+\a(u^2-1)$ 
was obtained and the amplitude $\a$ was estimated by four methods
described above: $\a_1$ from the least square estimator (eq. \ref{a_chi}),
$\a_3$ from the skewness (eq. \ref{a_3}), and $\a_5$ and $\a_7$ from the
fifth and seventh moments respectively 
(eq. \ref{a_5_7}).

In addition, the mean of four was also computed:
$\a_m=(\a_1+\a_3+\a_5+\a_7)/4$. 
Figures 6 and 7 show the probability functions $P(\a)$
measured in 400 realizations of the $256^2$ and $1024^2$ maps.
Four panels show results for four different amplitudes illustrating
the reliability of the detection of the non-Gaussianity and
measurement of the value of $\a$.
Five curves in each panel show the distribution functions 
$P(\a_1)$, $P(\a_3)$, $P(\a_5)$, $P(\a_7)$, and $P(\a_m)$. 
It is a little surprising that the curves 
heavily overlap which demonstrates that there is
no much difference in the distributions of $\a$ obtained by different
methods although the mean of four is always has a little shorter tails
(solid lines) while the estimate based on the highest moment $c_7$
often spreads a little wider than others (dotted lines). Although
all $\a_i$ are not independent they are the most sensitive to different
parts of the PF. The $\a_7$ is obviously sensitive to the higher values
of the PF than others and thus is probably the least reliable.

Comparing the theoretical estimates of the statistical limits
on the measurement of the quadratic non-Gaussianity (0.003 for the $256^2$ 
and 0.0008 for $1024^2$ map eq. \ref{limit_a}) with Fig. 6 and 7
we conclude that eq. \ref{limit_a} predicts the errors in measuring $\a$
quite accurately.

In these simulations the field $v$ obviously has a scale that
corresponds roughly to the pixel size. If the scale of the field was below
the resolution of the map the field would be smoothed. 
Smoothing obviously erases non-Gaussianity because of the central 
limit theorem.
We model the resolution effect by smoothing the original non-Gaussian
field with the Gaussian filter 
\begin{equation}
W(k,k_{sm}) = \exp (-\frac{k^2}{2k^2_{sm}}).
\end{equation} 

Real CMB maps are likely to be smoothed which obviously reduces the
ability of measuring $\alpha$. 
Figure 8 illustrates this effect. 
The quadratic non-Gaussianity can be easily detected and the amplitude 
$\a=0.01$ can be roughly estimated from a $256^2$ map if the scale 
of non-Gaussianity equals the pixel size.
In the top left panel the distribution function of $\alpha$ measured in
a thousand maps peaks at about right value  $\a=0.01$. The width
of the distribution function definitely allows the detection but
the accuracy of the measurement is not very high.  
If it was observed with four times lower resolution 
corresponding to the smoothing scale
$k_{sm}=64$ in the units of the fundamental wavenumber
the measured amplitude would be on average a half
of the original one. At the resolution 8 times of the original scale
of the non-Gaussianity the detection becomes impossible.
Of course, this depends on the amplitude. 
If the amplitude was  2.5 times higher ($\a=0.025$) then the detection 
of the non-Gaussianity would still possible
even at the resolution 8 times of the original scale as illustrated by 
by the right top panel of Fig. 9 however the measurement of the amplitude
would not be possible even it was ten times higher $\a=0.1$. Generally,
smoothing reduces non-Gaussianity and thus makes its measurement more
difficult.

\section{Summary}
The simplest tests for the primordial non-Gaussianity are based
on studies of the one-point functions: the PF or CPF  
and in particular the moments of the PF (skewness, kurtosis, \etc). 
However, the Gaussian form of the PF
does not guarantee the Gaussianity of the field. The hierarchy of the
n-point functions or n-spectra or other tests are supposed to 
analyze additional information that is not contained in the PF. 
It is natural to call it morphological information because it is
determined by the pattern of the contour lines only.
 
However, the sensitivity of n-point functions as well as many 
other statistics to the morphological information is unknown. 
In fact the n-point functions and other statistics use 
a great deal of the information stored in the PF. 
A simple example is provided by the analysis of the quadratic 
model (eq.\ref{GaG2}). 

At small amplitudes (practically at $\a < 0.1$ 
\footnote{One can see that even the value $\a_c =0.08$ 
requires a map with $N_{pix} > 10^9$ (Fig. 5) and eq. \ref{alpha_crit}
shows that $\a_c$ is extremely weak function of $N_{pix}$.}) 
the field becomes 
trivial non-Gaussian which means that the full information
about the non-Gaussianity is already contained in the PF 
(eq. \ref{pf_gag2_lim}). 

This means that nonzero three-point function
if applied to this model (eq. \ref{gag2_3pf})
uses exactly same statistical information as 
the PF. The morphological statistics in the form of the \mf of the
excursion set ($A$,  $C$ and $G$) and percolating region 
($A_p$, $C_p$ and $G_p$) obviously recycle the same information as the PF.
However, in the last case this information can be easily isolated in $A$
which is the CPF of the field. Other morphological statistics ($C$, $G$
$A_p$, $C_p$ and $G_p$) do not show any non-Gaussian signal if they
are expressed as functions of $A$ (Fig. 4). In general, these morphological
statistics use the different statistical information than the one-point
functions. 

Expressing the morphological
parameters as functions of $A$ instead of the level $u$ 
is equivalent to the gaussianization of
the field. For other (non morphological) statistics the gaussianization
of the field $v$ can be achieved by relabeling 
the levels of the field $v \rightarrow u$ according to eq. 
\ref{gausssianization}. 
The new field $u$ has the Gaussian PF by design but the geometry and   
topology of the original field. 
If any statistic sensitive to non-Gaussianity is
applied to the new field $u$ will characterize the geometry
and topology of the original field $v$ but free from the effects 
of the non-Gaussian PF of the original field. 
Performing this test  allows to assess
the level of independence of any statistic from the one-point
functions.

In line of the above  reasoning  we showed that the quadratic
model results in a trivial non-Gaussian field at small $\a$.
Using this fact we estimated the statistical limit
of detecting the quadratic model in the ideal situation when the
scale of the field is resolved. This limit is roughly $\a_{sl} \approx
N_{pix}^{-1/2}$ if the parent field $u$ is normalized to unity as
in eq. \ref{GaG2} or  $\a'_{sl}\sigma' \approx N_{pix}^{-1/2}$
if the rms of the parent field is $\sigma'$. The Monte Carlo simulations
on $256^2$ and $1024^2$ grids confirm this estimate. 

We measured the amplitude of the quadratic non-Gaussianity $\a$ using
four different estimators of $\a$: $\a_1$ from the least square fitting 
of the PF (eq. \ref{a_chi}) and from
three lowest order odd central moments $S=c_3, c_5$ and $c_7$
\begin{equation}  
\a_3=\frac{S}{6},~~ \a_5=\frac{c_5}{60},~~ \a_7=\frac{c_7}{630}.
\end{equation}
In addition, we also computed the mean of the four, 
$\a_m=(\a_1+\a_3+\a_5+\a_7)/4$.
Quite surprisingly, all four estimators $\a_1, \a_3, \a_5, \a_7$  
performed very similarly in all tests. This feature 
in combination  with the Gaussian
character of morphological statistics can be used for
distinguishing the quadratic non-Gaussianity from other types
of non-Gaussianity.
 
We have illustrated how the resolution and smoothing can affect 
the detection and measurement of the quadratic non-Gaussianity. 
In order to be detected the scale
of the original non-Gaussian field must be relatively close 
to the resolution of the map. However, quantifying this effect
requires the study of three-dimensional  parameter space 
(the size of the map, $N_{pix}$, the amplitude of the quadratic
non-Gaussianity, $\a$, and the resolution or smoothing scale, $k_{sm}$)
and is beyond the scope of this paper. 

\noindent {\bf Acknowledgments:}
I acknowledge the support of the GRF 2001 and 2002 grants at 
the University of Kansas. I am grateful to the referee for useful
comments and suggestions.

\newpage

\clearpage
\setcounter{figure}{0}
\begin{figure*}
\scalebox{0.4}{\includegraphics*[-3in,2in][8in,10in]{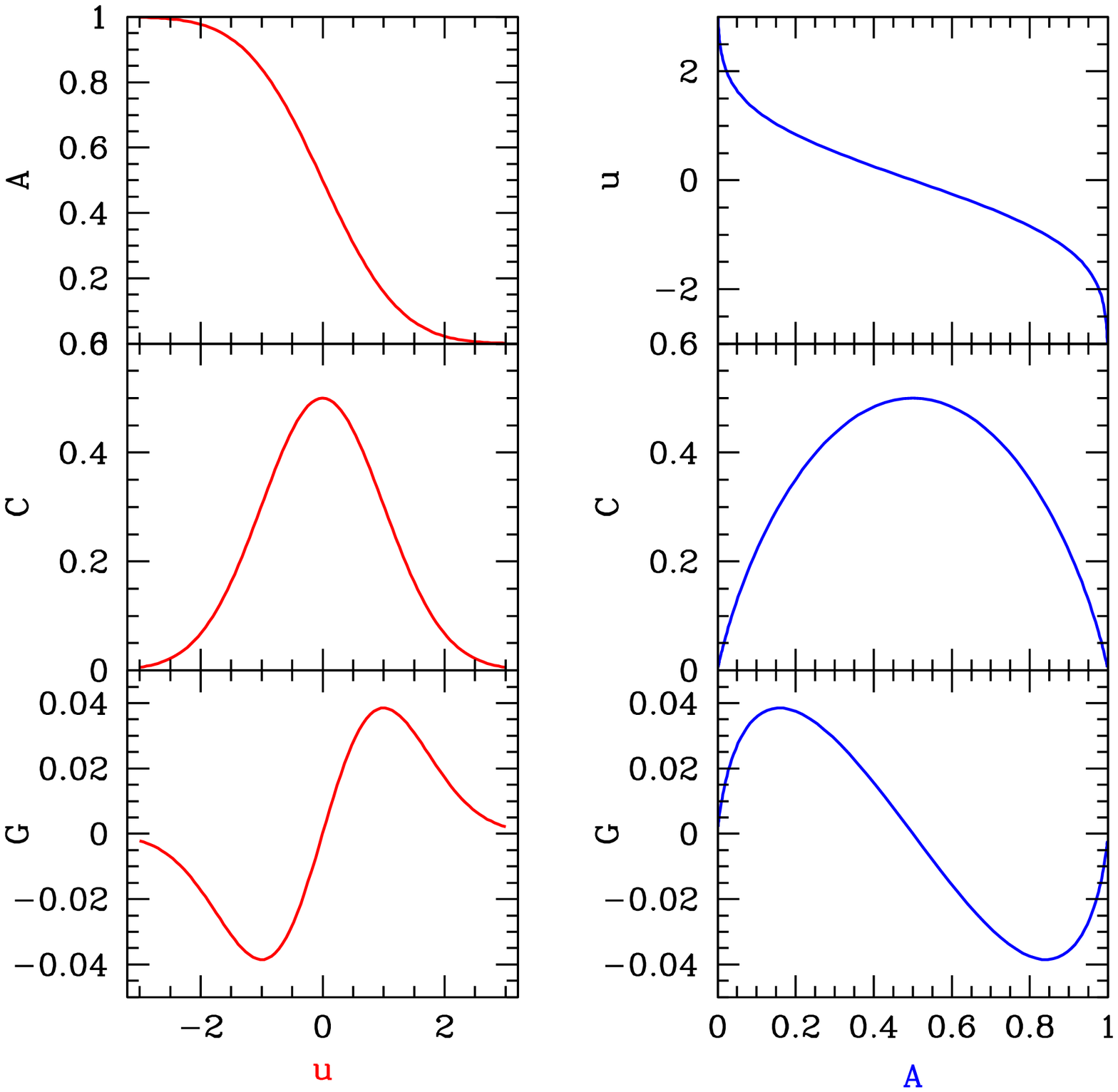}}
   \caption {Global Minkowski functionals of a Gaussian field. 
Three panels on the left hand-side show three 
global Minkowski functionals: the total fraction of the area in the 
excursion set, $A$, contour length per unit area, $C$, and
genus per unit area, $G$, as a function of the level, $u$. 
The panels on the right hand-side show (from top to bottom) 
the level, $u$, contour length, $C$, and genus $G$ as a function 
of the total area $A$.}
\label{fig1}
\end{figure*}

\begin{figure*}
\scalebox{0.4}{\includegraphics*[-3in,2in][8in,10in]{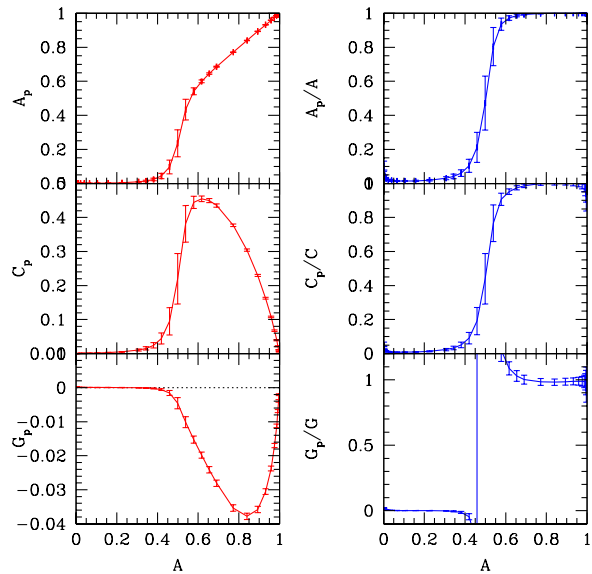}}
\caption {Percolation curves of Gaussian fields. 
Three panels on the left hand-side show: the area, $A_p$, 
contour length, $C_p$ and genus, $G_p$ of the largest region as 
a function of $A$ in the same units as in Fig.1. 
The right hand side panels show
the ratios of the \mf of the percolating region to the corresponding 
global \mf. The error bars show $1\sigma$ errors of the Monte Carlo
simulations}
\label{fig2}
\end{figure*}

\begin{figure*}
\scalebox{0.4}{\includegraphics*[-3in,2in][8in,10in]{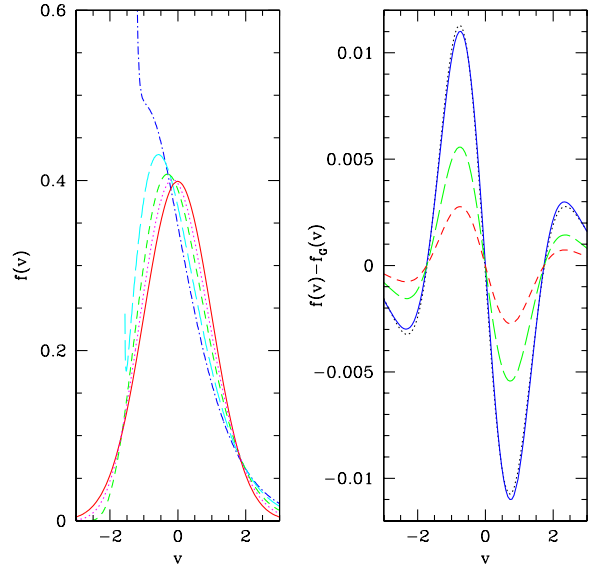}}
\caption {The probability functions in the quadratic model for 
five amplitudes: $\a=0.25, 0.18, 0.1$ and $0.05$ corresponding to
the dot-dashed, long dashed, short dashed, and dotted lines
are shown in the left hand side panel. The solid 
line is the Gaussian PF. The difference between 
the PF of the quadratic model and the Gaussian PF for three amplitudes
: $\a=0.02, 0.01$ and $0.005$ corresponding to dotted, long dashed 
and short dashed lines is shown if the right had side panel. 
The solid line shows the analytic linear approximation 
(eq. \ref{pf_gag2_lim}) with the amplitude $\a=0.02$.}
\label{fig3}
\end{figure*}

\begin{figure*}
\scalebox{0.4}{\includegraphics*[-3in,2in][8in,10in]{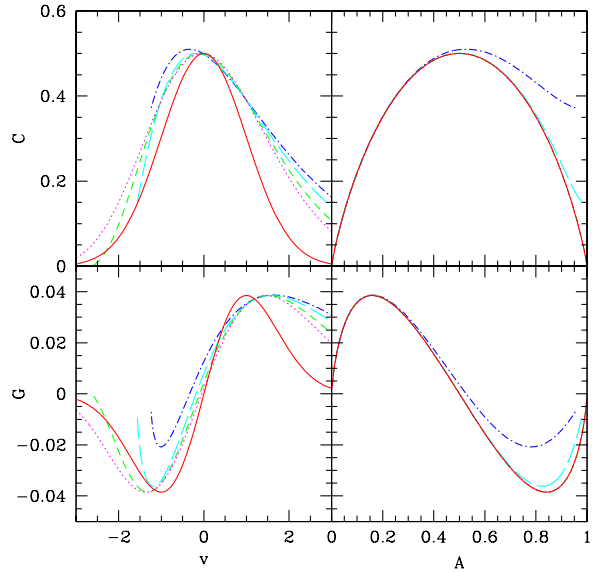}}
\caption {The Global Minkowski functionals $C$ and $G$ for the quadratic model 
parameterized by the level $v$ are shown in the left hand side panel.
the parameterization by the area 
of the excursion set $A$ is shown in the right hand side panels.
The amplitudes are as in the left
hand side panel of Fig. 3: $\a=0.25, 0.18, 0.1$ and $0.05$ corresponding to
the dot-dashed, long dashed, short dashed, and dotted lines. The solid 
lines show the Gaussian \mf. }
\label{fig4}
\end{figure*}

\begin{figure*}
\scalebox{0.4}{\includegraphics*[-3in,2in][8in,10in]{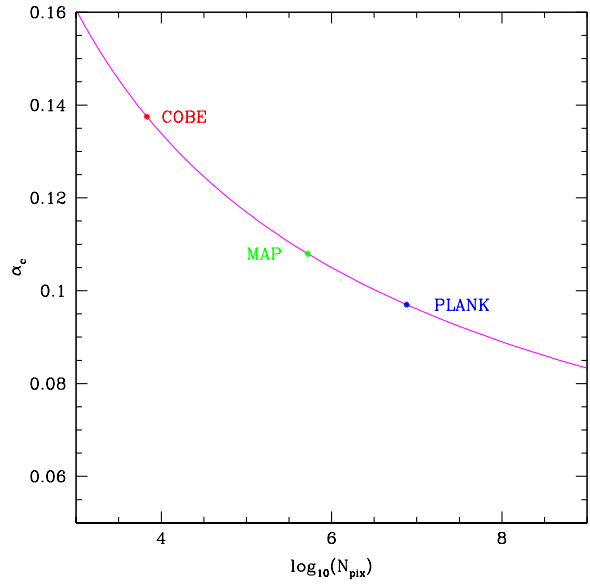}}
\caption {The triviality limit for the quadratic model $\a_c = \a_c(N_{pix})$.
If the amplitude $\a < \a_c$ then the quadratic model
is fully described by the PF only on a grid of the size $N_{pix}$.}
\label{fig5}
\end{figure*}

\begin{figure*}
\scalebox{0.4}{\includegraphics*[-3in,2in][8in,10in]{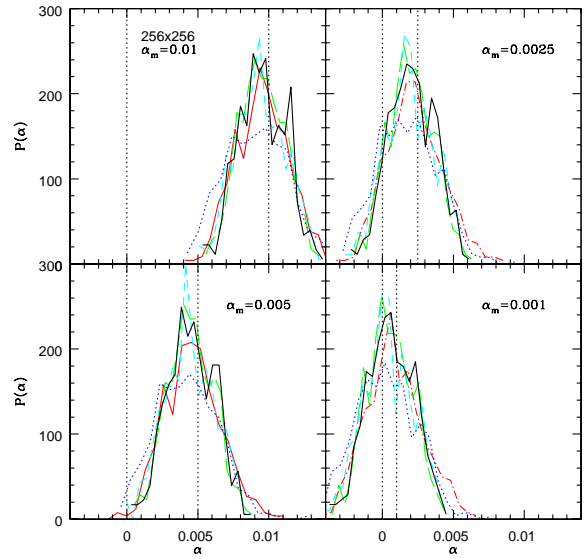}}
\caption {The probability functions of the amplitudes measured in
the Monte Carlo simulations of the $256^2$ maps. The amplitudes
$\a_1$ (obtaind from eq. \ref{a_chi}), $\a_3$ (eq. \ref{a_3}), $\a_5,\a_7$ 
 (eq. \ref{a_5_7}) and $\a_m =(\a_1+\a_3+\a_5+\a_7)/4$ are shown by the
dot-dashed, long dashed, short dashed, dotted, and solid lines respectively.
The vertical dotted lines mark $\a=0$ and the model value $\a_m$
in the simulations.}
\label{fig6}
\end{figure*}

\begin{figure*}
\scalebox{0.4}{\includegraphics*[-3in,2in][8in,10in]{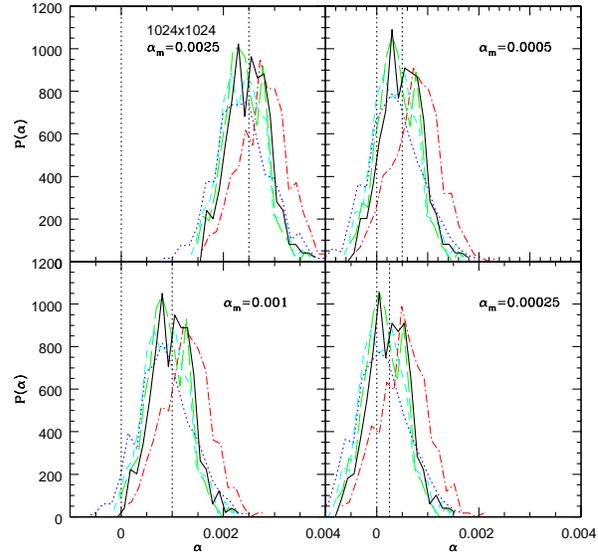}}
\caption {The probability functions of the amplitudes measured in
the Monte Carlo simulations of the $1024^2$ map. Notations are as in 
Fig. \ref{fig6}}
\label{fig7}
\end{figure*}

\begin{figure*}
\scalebox{0.4}{\includegraphics*[-3in,2in][8in,10in]{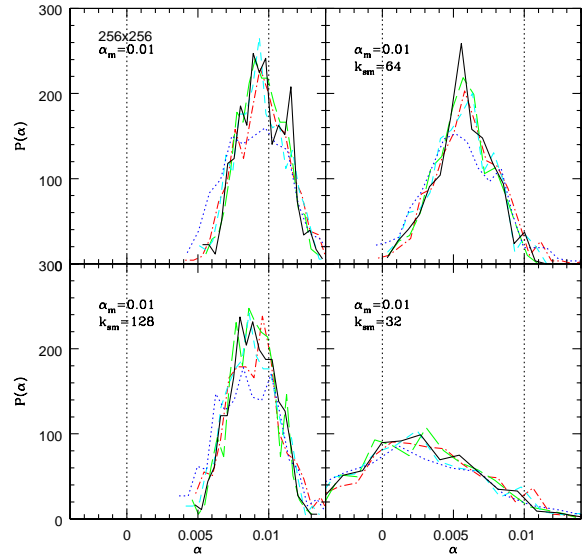}}
\caption {The effect of resolution in the $256^2$ maps. 
Four panels show the PFs
of the measured amplitudes when the initial quadratic field
was unsmoothed (left top), filtered with the Gaussian window
with the Nyquist frequency $k_{sm}=128$ (left bottom), 
with $k_{sm}=64$ (right top), and with $k_{sm}=32$ (right bottom).
Notations are as in Fig. \ref{fig6}}
\label{fig8}
\end{figure*}

\begin{figure*}
\scalebox{0.4}{\includegraphics*[-3in,2in][8in,10in]{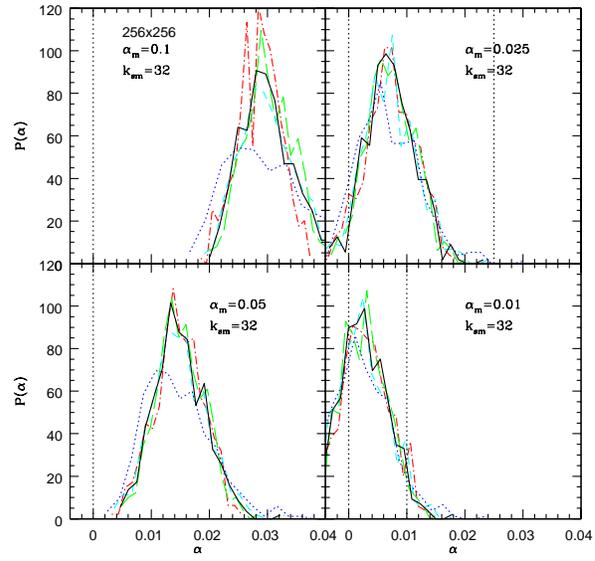}}
\caption {The effect of the amplitude in the $256^2$ maps. 
Four panels show the PFs
of the measured amplitudes if the initial quadratic field
smoothed with $k_{sm}=32$ had the amplitudes $\a=0.1$ (left top),
$\a=0.05$ (left bottom), $\a=0.025$ (right top), 
and  $\a=0.01$ (right bottom). Notations are as in Fig. \ref{fig6}}
\label{fig9}
\end{figure*}
\end{document}